\documentclass[review]{elsarticle}

\usepackage{amsthm,amsmath,amsfonts,amssymb}
\usepackage{subcaption}
\usepackage{lineno, hyperref}
\modulolinenumbers[5]
\usepackage{color}
\usepackage{ulem}
\usepackage[T1]{fontenc}
\usepackage[utf8]{inputenc}
\usepackage{float}
\usepackage{mathrsfs}

\newcommand{\vertiii}[1]{{\left\vert\kern-0.25ex\left\vert\kern-0.25ex\left\vert #1
\right\vert\kern-0.25ex\right\vert\kern-0.25ex\right\vert}}
\newcommand{\be}{\begin{equation}\begin{aligned}}
\newcommand{\ee}{\end{aligned}\end{equation}}
\newcommand{\ben}{\begin{equation}\nonumber\begin{aligned}}

\newcommand{\iu}{{i\mkern1mu}}

\def\e1{{\varepsilon_{11}}}
\def\b1{{\beta_{11}}}
\def\bp3{{\beta_{33}}}
\def\ep3{{\varepsilon_{33}}}

\begin{document}

\begin{frontmatter}	

\title{Numerical analysis of the Maxwell-Cattaneo-Vernotte nonlinear model}

\author[address1]{A. J. A. Ramos}
\ead{ramos@ufpa.br}

\author[address2]{A. D. S. Campelo}
\ead{campelo@ufpa.br}

\author[address1]{M. M. Freitas}
\ead{mirelson@ufpa.br}

\author[address3,address4]{R. Kovács\corref{mycorrespondingauthor}}
\ead{kovacs.robert@wigner.hu}

\address[address1]{Faculty of Mathematics, Federal University of Par\'a, Raimundo Santana Street s/n, Salin\'opolis--PA, 68721-000, Brazil}

\address[address2]{PhD Program in Mathematics, Federal University of  Par\'a, Augusto Corr\^ea Street 01,
Bel\'em--PA, 66075-110, Brazil}

\address[address3]{Department of Energy Engineering, Faculty of Mechanical Engineering, Budapest University of Technology and Economics, Muegyetem rkp. 3., H-1111 Budapest, Hungary}

\address[address4]{Department of Theoretical Physics, Wigner Research Centre for Physics, Budapest, Hungary}

\begin{abstract}
In the literature, one can find numerous modifications of Fourier’s law from which the first one is called Maxwell-Cattaneo-Vernotte heat equation. Although this model has been known for decades and successfully used to model low-temperature damped heat wave propagation, its nonlinear properties are rarely investigated. In this paper, we aim to present the functional relationship between the transport coefficients and the consequences of their temperature dependence. Furthermore, we introduce a particular implicit numerical scheme in order to solve such nonlinear heat equations reliably. We investigate the scheme’s stability, dissipation, and dispersion attributes as well. We demonstrate the effect of temperature-dependent thermal conductivity on two different initial-boundary value problem, including time-dependent boundaries and heterogeneous initial conditions.

\end{abstract}
\vspace{0.5cm}

\end{frontmatter}

\section{Introduction}

In recent years, numerous heat conduction models have been developed to provide a more efficient modeling tool for complex problems related to wave propagation under low-temperature conditions \cite{JacWalMcN70, Don09}, in rarefied media \cite{Struc05, Arietal15, RhaStruc16}, in nanosystems \cite{SellEtal16b, Chen00}, or over-diffusion in complex heterogeneous material structures \cite{FehEtal21, LunEtal22}. The basic properties of the heat equations depend on the particular thermodynamic background. For instance, the approach of Rational Extended Thermodynamics (RET) \cite{MulRug98, RugSug15} exploits kinetic theory rigorously, thus it requires particular assumptions about the microscopic mechanisms, and resulting in a model with given transport coefficients. A continuum theory, on the contrary, does not need any prior assumption, and it remains arbitrary whether the continuum equations inherit the particular coefficients from RET. Such approaches are called Extended Irreversible Thermodynamics (EIT) \cite{JouEtal10b, Lebon14} and Non-Equilibrium Thermodynamics with Internal Variables (NET-IV) \cite{Gyarmati70b, Verhas96, BerVan17b}. Furthermore, while RET derives the balances through a momentum series expansion of the Boltzmann transport equation, EIT and NET-IV starts with the balances and derives the constitutive equation from the second law of thermodynamics, using Onsagerian relations. These procedures are discussed in detail in \cite{VanFul12}. Here, we want to focus on the simplest heat equation beyond Fourier, called Maxwell-Cattaneo-Vernotte (MCV) or Cattaneo equation \cite{Cattaneo58, Vernotte58}, following from a continuum theory. Hence, it reads
\begin{align}
\tau q_{t}+q=-\lambda T_{x},
\end{align}
where $\tau$ and $\lambda$ are the relaxation time and thermal conductivity coefficients, which are not related to any microscopic mechanism. Furthermore, for our purpose, a one-dimensional rigid, isotropic conductor is satisfactory, in which $q$ and $T$ are the heat flux and temperature. In the indices, $t$ and $x$ stand for the corresponding partial derivative. Although the validity of the MCV equation is restricted to low-temperature situations \cite{Auriault16}, e.g., it can model a damped wave propagation (second sound) well, and thus its role remain marginal in standard engineering practice, there are crucial properties need to be understood. 

We place the emphasis on nonlinearities, particularly, on the temperature-dependent coefficients. In the previous work of Rogolino and Kovacs \cite{KovRog20}, based on the Onsagerian form of the MCV equation,
\begin{align}
\left (\frac{1}{T}\right)_x - \rho(T) m\, q_t -l(T) q=0, \quad m>0, \quad l>0, \label{GKE}
\end{align}
it has been underlined that the coefficients are not independent of each other. For instance, assuming a linear $T$-dependent thermal conductivity,
\begin{align}
\hat \lambda = \lambda_{0}+a(T-T_{0}), \quad \lambda_{0} = \hat \lambda(T_0) >0, \quad a\in\mathbb{R},
\end{align}
influences the relaxation time as well. Since $m>0$ is a constant (otherwise further terms would enter the constitutive equation \eqref{GKE}), the mass density $\rho$ must depend on the temperature as well in order to achieve the desired $\tau=\tau(T)$ dependence due to $l(T)$
\begin{align}
\frac{\rho(T) m}{l(T)} q_t +q=-\frac{1}{l(T)}\frac{1}{T^2}T_x, \quad \tau(T)=\frac{\rho(T) m}{l(T)}, \quad \lambda(T)= \frac{1}{l(T)}\frac{1}{T^2}. \label{GKE3}
\end{align}
Consequently,
\begin{align}
l(T)=\frac{1}{\lambda_{0}+a(T-T_{0})T^{2}} \qquad  \mbox{and} \qquad \rho(T)=\frac{\tau}{m} l(T),
\end{align}
and due to $\rho(T)$, mechanics should be involved into the modeling. In other words, despite having the simplest extension of Fourier's law, and adding a straightforward temperature dependence in the thermal conductivity, leads to a complicated thermo-mechanical model. 
Furthermore, in \cite{KovRog20}, an explicit finite difference technique is utilized for which one needs to determine the stability criteria, and the numerical solutions are suffered from artificial dispersion error.

Since such model is nonlinear, the stability properties are not straightforward to determine. In the present paper, we want to provide further insights into the numerical solution of the nonlinear MCV equation, however, for our purposes, we need to simplify Eq.~\eqref{GKE3}. Although any simplification inevitably truncate the physical content of this nonlinear model, it is satisfactory to keep the relaxation time $\tau$ to be constant, and we highlight that it limits the physical validity of the model.
This yields
\begin{align}
\tau q_t +q+\big[\lambda_{0}+a(T-T_{0})\big]T_x =0. 
\end{align}
After some manipulations we have
\begin{align}
\tau q_{t}+q+aT\,T_{x}+\lambda T_{x}=0, \label{eq1}
\end{align}
where $\lambda=\lambda_{0}-aT_{0}$.
In the present paper, we limit ourselves on Eq.~\eqref{eq1}, and introduce an implicit numerical approach to efficiently handle the nonlinear term $aT T_x$. We prove that the implicit discretization is unconditionally stable, thus it is now free from stability issues. Additionally, we also prove that the numerical solution has minimal distortion by dissipation and it is also free from dispersion errors.

\section{Initial and boundary conditions}
In the following, we consider the MCV model in the form
\begin{align}
\rho c T_{t}+q_{x}&=0 \quad \mbox{in} \quad (0,\ell) \times (0,\infty), \label{Eq1}
\\
\tau q_{t}+q+aT\,T_{x}+\lambda T_{x}&=0 \quad \mbox{in} \quad (0,\ell) \times (0,\infty), \label{Eq2}
\end{align}
where Eq.~\eqref{Eq1} supplements the constitutive equation, called balance equation of internal energy $e$, for which we assumed that $e=cT$ with $c$ being the specific heat, and all volumetric heat sources are omitted. The length of the conducting medium is $\ell$.
We consider two types of boundary conditions:
\begin{eqnarray}
&&\mbox{\textbf{Boundary type I:}} \quad \begin{cases}
q(0,t)=0, \quad \mbox{for all} \quad t\geq0,
\\
q(\ell,t)=0, \quad \mbox{for all} \quad t\geq0,\label{CCont1a}
\end{cases}
\\	
\nonumber\\	
\nonumber\\	
&&\mbox{\textbf{Boundary type II:}} \hspace{-0.12cm}\quad \begin{cases}
q(0,t)=\begin{cases}
1-\cos\big(2\pi t/t_{p}\big), \quad \mbox{if} \quad 0<t\leq t_{p}, \quad t_{p}>0
\\
0, \quad \mbox{if} \quad t> t_{p},\label{CCont2a}
\end{cases}
\\
\\
q(\ell,t)= 0, \quad \mbox{for all} \quad t\geq0,
\end{cases}
\end{eqnarray}
for which we assign also two types of initial conditions:
\begin{eqnarray}
\mbox{\textbf{Initial condition I:}} \quad T(x,0)=T_{0}(x), \quad q(x,0)=q_{0}(x),\quad x \in (0,\ell), \label{CInitial}
\end{eqnarray}
\begin{eqnarray}
\mbox{\textbf{Initial condition II:}} \quad T(x,0)=T_{0}, \quad q(x,0)=q_{0}\equiv0,\quad x \in (0,\ell). \label{CInitial2}
\end{eqnarray}
The type I.~initial and boundary conditions represent a situation of heterogeneous initial state, the spatial-dependent temperature distribution induce a non-homogeneous heat flux field. We observe the time evolution of such system with adiabatic boundary conditions, therefore the resulting temperature distribution is not influenced by environmental conditions such as heat convection.

The type II.~setting, however, displays the usual conditions of a heat pulse experiment, the initial steady-state is excited by a heat pulse with duration of $t_p$. The present particular form of Eq.~\eqref{CCont2a} is advantageous from a numerical point of view. The heat flux boundary is initiated with zero derivative, therefore artificial oscillations of such source can be avoided.

\section{Numerical linearization method }

Let us rewrite the system of equations (\ref{Eq1})--(\ref{Eq2}) in the form
\begin{align}
\rho c T_{t}+q_{x}&=0 \quad \mbox{in} \quad (0,\ell) \times (0,\mathcal{T}), \label{Eq1b}
\\
\tau q_{t}+q+\frac{a}{2}\big(T^{2}\big)_{x}+\lambda T_{x}&=0 \quad \mbox{in} \quad (0,\ell) \times (0,\mathcal{T}), \label{Eq2b}
\end{align}
and use an implicit finite difference method to discretize the system (\ref{Eq1b})--(\ref{Eq2b}). More precisely, we consider $ J,N \in \mathbb{N}$, and set $ \Delta x=\displaystyle \frac{\ell}{J+1},  \Delta t=\displaystyle \frac{\mathcal{T}}{N+1} $ and we introduce a uniform mesh
\begin{eqnarray}
&&0=x_{0}<x_{1} < \cdot\cdot\cdot< x_{j}=j\Delta x<\cdot\cdot\cdot<x_{J}<x_{J+1}=\ell, \quad j=0,1,..., J+1,
\\
&&0=t_{0}<t_{1} < \cdot\cdot\cdot< t_{n}=n\Delta t<\cdot\cdot\cdot<t_{N}<t_{N+1}=\mathcal{T}, \quad n=0,1,..., N+1.
\end{eqnarray}
where $\mathcal{T}$ denotes the entire time interval used in the simulations, as well as the indices $j$ and $n$ stand for the corresponding space and time steps. 
We construct the implicit numerical scheme as
\begin{eqnarray}
&&\rho c \frac{T_{j,n}-T_{j,n-1}}{\Delta t}+\frac{q_{j+1,n}-q_{j,n}}{\Delta x} =0,\quad
j=0,1,...,J, \quad  n=1,2,...,N,\label{eq.MCV.01}
\\
&&\tau\frac{q_{j,n}-q_{j,n-1}}{\Delta t}+q_{j,n}+\frac{a}{2}\frac{T_{j,n}^{2}-T_{j-1,n}^{2}}{\Delta x}+\lambda\frac{T_{j,n}-T_{j-1,n}}{\Delta x}=0, \quad
j=1,2,...,J, \quad  n=1,2,...,N,\label{eq.MCV.02}
\end{eqnarray}
supplemented with the discrete boundary,
\begin{eqnarray}
&&\mbox{\textbf{Boundary type I:}} \quad \begin{cases}
q_{0,n}=0, \quad \mbox{for all} \quad n=0,1,...,N+1, 
\\
q_{J+1,n}=0, \quad \mbox{for all} \quad n=0,1,...,N+1,\label{CCont1c}
\end{cases}
\\	
\nonumber\\	
\nonumber\\	
&&\mbox{\textbf{Boundary type II:}} \hspace{-0.12cm}\quad\begin{cases}
q_{0,n}=\begin{cases}
1-\cos\big(2\pi t_{n}/t_{p}\big), \quad \mbox{if} \quad 0<n\leq p, \quad p\in \mathbb{N},
\\
0, \quad \mbox{if} \quad n>p,\label{CCont2d}
\end{cases}
\\
\\
q_{J+1,n}= 0, \quad \mbox{for all} \quad n=0,1,..., N+1,
\end{cases}
\end{eqnarray}
and initial conditions,
\begin{eqnarray}\label{eq.MCV.04}
T_{j,0}=T_{j}^{0}, \quad q_{j,0}=q_{j}^{0}, \quad \mbox{for all} \quad j=0,1,...,J+1.
\end{eqnarray}
The scheme, in its present state, is also nonlinear due to the terms $T_{j,n}^2$, corresponding to the unknown temperature value at the new time instant. However, while preserving that nonlinearity in \eqref{eq.MCV.02}, we can devolve it from $n$ to $n-1$. Let us assume that $T(x, t)$ is sufficiently regular and use the Taylor expansion to write
\begin{eqnarray*}
T^{2}(x, t-\Delta t)=T^{2}(x, t)-\Delta t \frac{\partial T^{2}(x, t)}{\partial T}\frac{\partial T}{\partial t}+O(\Delta t^{2}).
\end{eqnarray*}
Using backward difference to approximate $T_t$, it yields
\begin{eqnarray*}\label{eq.y2a}
T^{2}(x, t-\Delta t)=T^{2}(x, t)-2T(x, t)\big(T(x,t)-T(x,t-\Delta t)\big)+O(\Delta t^{2}).
\end{eqnarray*}
Furthermore, 
\begin{eqnarray}\label{eq.y2b}
\nonumber T^{2}(x, t)&=&-T^{2}(x, t-\Delta t)+2T(x, t)T(x, t-\Delta t)+O(\Delta t^{2})
\\
&=&T^{2}(x, t-\Delta t)+2T(x, t-\Delta t)\big(T(x, t)-T(x, t-\Delta t)\big)+O(\Delta t^{2})
\end{eqnarray}
holds. Now we use the Taylor expansion given in (\ref{eq.y2b}) to obtain an approximation for the nonlinear term $(T_{j,n}^{2}-T_{j-1,n}^{2 })/\Delta x$ given in (\ref{eq.MCV.02}), that is, we consider
\begin{eqnarray}\label{approx}
\frac{T_{j,n}^{2}-T_{j-1,n}^{2}}{\Delta x}&\approx&\frac{T_{j,n-1}^{2}-T_{j-1,n-1}^{2}}{\Delta x}+2\frac{T_{j,n-1}\big(T_{j,n}-T_{j,n-1}\big)-T_{j-1,n-1}\big(T_{j-1,n}-T_{j-1,n-1}\big)}{\Delta x}.
\end{eqnarray}
Substituting $(\ref{approx})$ into (\ref{eq.MCV.02}),
\begin{eqnarray}
&&\rho c \frac{T_{j,n}-T_{j,n-1}}{\Delta t}+\frac{q_{j+1,n}-q_{j,n}}{\Delta x} =0,\quad
j=0,1,...,J, \quad  n=1,2,...,N,\label{eq.aprox.01}
\\
\nonumber&&\tau\frac{q_{j,n}-q_{j,n-1}}{\Delta t}+q_{j,n}+\frac{a}{2}\frac{T_{j,n-1}^{2}-T_{j-1,n-1}^{2}}{\Delta x}+a\frac{T_{j,n-1}\big(T_{j,n}-T_{j,n-1}\big)-T_{j-1,n-1}\big(T_{j-1,n}-T_{j-1,n-1}\big)}{\Delta x}
\\
&&+\lambda\frac{T_{j,n}-T_{j-1,n}}{\Delta x}=0,\quad
j=1,2,...,J, \quad  n=1,2,...,N.\label{eq.aprox.02}
\end{eqnarray}
so that the nonlinearity does not affect the calculation of the new time instants, this is a sort of numerical linearization we performed.


For any $n$, we can introduce $\phi_{j}:=T_{j,n}-T_{j,n-1}$ and $\psi_{j}:=q_{j,n}-q_{j ,n-1}$, obtaining the system
\begin{eqnarray}
&&\rho c \phi_{j}+r(\psi_{j+1}-\psi_{j}) =-r(q_{j+1,n-1}-q_{j,n-1}), \quad j=0,1,...,J,\label{eq.MCV.01b}
\\
\nonumber&&2(\tau+\Delta t)\psi_{j}+2r(\lambda+aT_{j,n-1})\phi_{j}-2r(\lambda+aT_{j-1,n-1})\phi_{j-1}=-ar(T_{j,n-1}^{2}-T_{j-1,n-1}^{2})
\\
&&-2\lambda r(T_{j,n-1}-T_{j-1,n-1})-2\Delta tq_{j,n-1}\quad j=1,2,...,J,
\label{eq.MCV.02b}
\end{eqnarray}
for $\{\phi_{0}, \phi_{1},...,\phi_{J}\}$ and $\{\psi_{1}, \psi_{2},...,\psi_{J}\}$, with $r=\Delta t/\Delta x$. In this case, the representation of type I.~and type II.~boundary conditions are given by
\begin{eqnarray}
&&\mbox{\textbf{Boundary type I:}} \quad \begin{cases}
\psi_{0}=0, 
\\
\psi_{J+1}=0,\label{CCont1e}
\end{cases}
\\	
\nonumber\\	
\nonumber\\	
&&\mbox{\textbf{Boundary type II:}} \hspace{-0.12cm}\quad\begin{cases}
\psi_{0}=\begin{cases}
\cos\big(2\pi t_{n-1}/t_{p}\big)-\cos\big(2\pi t_{n}/t_{p}\big), \quad \mbox{if} \quad 0<n\leq p, \quad p\in \mathbb{N},
\\
0, \quad \mbox{if} \quad n>p,\label{CCont1f}
\end{cases}
\\
\\
\psi_{J+1}= 0.
\end{cases}
\end{eqnarray}
This is more suitable to rewrite the difference equations into a matrix form, therefore, let us rewrite the scheme (\ref{eq.MCV.01b})--(\ref{eq.MCV.02b}) in an equivalent vectorial form, using the matrices
$$
\textbf{A}:=\left(\begin{array}{cccccc}
1&  0&  0&  0&  \cdots& 0  \\
-1&  1&  0&  \ddots&  \ddots& \colon  \\
0&  -1&  \ddots&  \ddots&  \ddots& 0  \\
0&  \ddots&  \ddots&  \ddots&  0& 0  \\
\colon&  \ddots&  \ddots&  -1&  1& 0 \\
0&  \cdots&  0&  0&  -1& 1 \\
0&  \cdots&  0&  0&  0& -1   
\end{array}\right)_{J+1\times J}, \ \ \textbf{B}:=\left(\begin{array}{ccccccc}
b_{0}&  c_{1}&  0&  0&  \cdots& 0  & 0 \\
0&  b_{1}&  c_{2}&  \ddots&  \ddots& \colon  & 0   \\
0&  0&  \ddots&  \ddots&  \ddots& 0  & 0   \\
0&  \ddots&  \ddots&  \ddots&  c_{1}& 0  & 0   \\
\colon&  \ddots&  \ddots&  0&  b_{J-2}& c_{J-1}  & 0  \\
0&  \cdots&  0&  0&  0& b_{J-1}  & c_{J}    
\end{array}\right)_{J\times J+1}
$$
and
$$
\textbf{C}:=\left(\begin{array}{ccccccc}
-1&  1&  0&  0&  \cdots& 0  & 0 \\
0&  -1&  1&  \ddots&  \ddots& \colon  & 0   \\
0&  0&  \ddots&  \ddots&  \ddots& 0  & 0   \\
0&  \ddots&  \ddots&  \ddots&  1& 0  & 0   \\
\colon&  \ddots&  \ddots&  0&  -1& 1  & 0  \\
0&  \cdots&  0&  0&  0& -1  & 1    
\end{array}\right)_{J\times J+1} \hspace{-0.5cm}, \  \textbf{D}:=\left(\begin{array}{cccccc}
T_{0, n-1}	&  0&  0&  0&  \cdots& 0  \\
0&  T_{1, n-1}&  0&  \ddots&  \ddots& \colon  \\
0&  0&  \ddots&  \ddots&  \ddots& 0  \\
0&  \ddots&  \ddots&  \ddots&  0& 0  \\
\colon&  \ddots&  \ddots&  0&  T_{J-1, n-1}& 0 \\
0&  \cdots&  0&  0&  0& T_{J, n-1}  
\end{array}\right)_{J+1\times J+1},
$$
where $b_{j-1}=-2r(\lambda+aT_{j-1,n-1})$ and $c_{j}=2r(\lambda+aT_{j,n-1})$, hence $\Phi=(\phi_{0}, \phi_{1},...,\phi_{J})^{\top}$, $\Psi=(\psi_{1}, \psi_{2},...,\psi_{J})^{\top}$, $\mathbb{T}^{n-1}=(T_{0, n-1}, T_{1, n-1},...,T_{J, n-1})^{\top}$ and $\mathbb{Q}^{n-1}=(q_{1, n-1}, q_{2, n-1},...,q_{J, n-1})^{\top}$. 

\subsubsection*{Boundary type I.} The scheme (\ref{eq.MCV.01b})--(\ref{eq.MCV.02b}) with boundary type I.~takes the following vector form:
\begin{eqnarray}\label{numerical.problem1}
\begin{cases}
\displaystyle \Phi+\frac{r}{\rho c}\textbf{A}\Psi=-\frac{r}{\rho c}\textbf{A}\mathbb{Q}^{n-1}, 
\\
\displaystyle \Psi+\frac{1}{2\big(\tau+\Delta t\big)}\textbf{B}\Phi=-\frac{r}{2\big(\tau+\Delta t\big)}\textbf{C}\Big(a\textbf{D}+2\lambda \textbf{I}_{J+1}\Big)\mathbb{T}^{n-1}-\frac{\Delta t}{\big(\tau+\Delta t\big)}\mathbb{Q}^{n-1},
\end{cases}
\end{eqnarray}
where $\textbf{I}_{J+1}$ is an identity matrix of order $J+1$. Combining the above equations, we obtain
\begin{eqnarray}
\begin{cases}
\displaystyle \Phi=\frac{r}{\rho c}\textbf{A}\bigg(\frac{r}{2(\tau+\Delta t)}\textbf{G}-\frac{1}{\tau+\Delta t}\textbf{F}-\mathbb{Q}^{n-1}\bigg)
\\
\displaystyle \Psi=\frac{1}{\tau+\Delta t}\textbf{F}-\frac{r}{2(\tau+\Delta t)}\textbf{G},
\end{cases}
\end{eqnarray}
in which 
\begin{eqnarray}
\textbf{E}:=\textbf{I}_{J}-\frac{r}{2\rho c(\tau+\Delta t)}\textbf{B}\textbf{A}, \quad \textbf{F}:=\textbf{E}^{-1}\bigg(\frac{r}{2\rho c}\textbf{B}\textbf{A}-\Delta t\,\textbf{I}_{J}\bigg)\mathbb{Q}^{n-1}, \quad
\textbf{G}:=\textbf{E}^{-1}\textbf{C}\bigg(a\textbf{D}+2\lambda\textbf{I}_{J+1}\bigg)\mathbb{T}^{n-1}
\end{eqnarray}
and $\textbf{I}_{J}$ is an identity matrix of order $J$. Finally, the solution of the numerical scheme (\ref{eq.MCV.01})--(\ref{eq.MCV.04}) is given by
\begin{eqnarray}\label{syst.01}
\begin{cases}
\displaystyle \mathbb{T}^{n}=\mathbb{T}^{n-1}+\Phi, \quad n=1,2,...,N,
\\
\displaystyle \mathbb{Q}^{n}=\mathbb{Q}^{n-1}+\Psi, \quad n=1,2,...,N,
\\
\displaystyle \mathbb{T}^{0}=(T_{0}^{0}, T_{1}^{0},...,T_{J}^{0})^{\top},  \quad \mathbb{Q}^{0}=(q_{1}^{0}, q_{2}^{0},...,q_{J}^{0})^{\top}.
\end{cases}
\end{eqnarray}

\subsubsection*{Boundary type II.}
The scheme (\ref{eq.MCV.01b})--(\ref{eq.MCV.02b}) with boundary type II.~takes the following vector form:
\begin{eqnarray}\label{numerical.problem2}
\begin{cases}
\displaystyle \Phi+\frac{r}{\rho c}\textbf{A}\Psi={
\begin{cases}
-\frac{r}{\rho c}\textbf{A}\mathbb{Q}^{n-1}+\frac{r}{\rho c}\Big(1-\cos\big(2\pi t_{n}/t_{p}\big)\Big)\textbf{L}, \quad \mbox{if} \quad 0<n\leq p, \quad p\in \mathbb{N} 
\\
-\frac{r}{\rho c}\textbf{A}\mathbb{Q}^{n-1}, \quad \mbox{if} \quad n>p,
\end{cases}}
\\
\\
\displaystyle \Psi+\frac{1}{2\big(\tau+\Delta t\big)}\textbf{B}\Phi=-\frac{r}{2\big(\tau+\Delta t\big)}\textbf{C}\Big(a\textbf{D}+2\lambda \textbf{I}_{J+1}\Big)\mathbb{T}^{n-1}-\frac{\Delta t}{\big(\tau+\Delta t\big)}\mathbb{Q}^{n-1},
\end{cases}
\end{eqnarray}
where $\textbf{L}=(1,0, \cdot\cdot\cdot,0)_{1 \times J+1}^{\top}$ and $\textbf{I}_{J+1}$ is an identity matrix of order $J+1$. Combining the above equations we obtain
\begin{eqnarray}
\begin{cases}
\displaystyle \Phi=
{
\begin{cases}
\frac{r}{\rho c}\textbf{A}\bigg(\frac{r}{2(\tau+\Delta t)}\textbf{G}-\frac{1}{\tau+\Delta t}\textbf{F}-\mathbb{Q}^{n-1}\bigg)+\frac{r}{\rho c}\Big(1-\cos\big(2\pi t_{n}/t_{p}\big)\Big)\textbf{L}
\\
\quad \ \ \displaystyle+\frac{r^{2}}{2\rho^{2}c^{2}(\tau+\Delta t)}\Big(1-\cos\big(2\pi t_{n}/t_{p}\big)\Big)\textbf{A}\textbf{E}^{-1}\textbf{B}\textbf{L}, \quad \mbox{if} \quad 0<n\leq p, \quad p\in \mathbb{N} 
\\
\\
\frac{r}{\rho c}\textbf{A}\bigg(\frac{r}{2(\tau+\Delta t)}\textbf{G}-\frac{1}{\tau+\Delta t}\textbf{F}-\mathbb{Q}^{n-1}\bigg), \quad \mbox{if} \quad n>p,
\end{cases}}
\\
\\
\displaystyle \Psi=
{
\begin{cases}
\frac{1}{\tau+\Delta t}\textbf{F}-\frac{r}{2(\tau+\Delta t)}\textbf{G}-\frac{r}{2\rho c(\tau+\Delta t)}\Big(1-\cos\big(2\pi t_{n}/t_{p}\big)\Big)\textbf{E}^{-1}\textbf{B}\textbf{L}, \quad \mbox{if} \quad 0<n\leq p, \quad p\in \mathbb{N} 
\\
\\
\frac{1}{\tau+\Delta t}\textbf{F}-\frac{r}{2(\tau+\Delta t)}\textbf{G}, \quad \mbox{if} \quad n>p,
\end{cases}}
\end{cases}
\end{eqnarray}
with
\begin{eqnarray}
\textbf{E}:=\textbf{I}_{J}-\frac{r}{2\rho c(\tau+\Delta t)}\textbf{B}\textbf{A}, \quad \textbf{F}:=\textbf{E}^{-1}\bigg(\frac{r}{2\rho c}\textbf{B}\textbf{A}-\Delta t\,\textbf{I}_{J}\bigg)\mathbb{Q}^{n-1}, \quad
\textbf{G}:=\textbf{E}^{-1}\textbf{C}\bigg(a\textbf{D}+2\lambda\textbf{I}_{J+1}\bigg)\mathbb{T}^{n-1},
\end{eqnarray}
and $\textbf{I}_{J}$ is an identity matrix of order $J$. Finally, the solution of the numerical scheme (\ref{eq.MCV.01})--(\ref{eq.MCV.04}) is given by
\begin{eqnarray}
\begin{cases}
\displaystyle \mathbb{T}^{n}=\mathbb{T}^{n-1}+\Phi, \quad n=1,2,...,N,
\\
\displaystyle \mathbb{Q}^{n}=\mathbb{Q}^{n-1}+\Psi, \quad n=1,2,...,N,
\\
\displaystyle \mathbb{T}^{0}=(T_{0}^{0}, T_{1}^{0},...,T_{J}^{0})^{\top},  \quad \mathbb{Q}^{0}=(q_{1}^{0}, q_{2}^{0},...,q_{J}^{0})^{\top}.
\end{cases}
\end{eqnarray}

\section{Stability, dissipation and dispersion}
Let us investigate the scheme \eqref{eq.aprox.01}-\eqref{eq.aprox.02}, using the conventional Neumann method \cite{Press07b}. Although it is developed for linear equations, it is still of good use for such a nonlinear situation since only the known values of the temperature field are nonlinear, not the new, hence unknown ones.
Following this procedure, we assume that 
\begin{align}
T_{j,n} = T_0 \xi^n e^{\iu k j \Delta x}, \quad q_{j,n} = q_0 \xi^n e^{\iu k j \Delta x}, \label{neu1}
\end{align}
where $T_0$ and $q_0$ are the initial amplitudes of the corresponding field quantity, $\iu$, $k$ and $\xi$ are the imaginary unit, wave number, and the wave amplitude, respectively. It is clear that to achieve a stable numerical solution, one needs $|\xi| \leq 1$, otherwise, the amplitude will grow up without limit.
Substituting Eq.~\eqref{neu1} into \eqref{eq.aprox.01} and \eqref{eq.aprox.02} is not a linearization, it results in a nonlinear algebraic equation for $\xi$. The substitution yields
\begin{align}
T_0 \frac{\rho c}{\Delta t} \big( 1- \xi^{-1}\big) + q_0 \frac{1}{\Delta x} \big(e^{\iu k \Delta x}-1\big) = 0, \label{neu2}
\end{align}
\begin{align}
q_0 \left(\frac{\tau}{\Delta t} \big(1- \xi^{-1}\big) +1 \right) + T_0 \frac{a}{\Delta x}\left(  \frac{1}{2} \xi^{n-1} e^{\iu k j \Delta x}\big(1-e^{-2\iu k \Delta x}   \big) + \big(\xi^{n-1} - \xi^{n-2}\big)\big(1 - e^{-\iu k \Delta x}\big) + \frac{\lambda}{a} \big(1 - e^{-\iu k  \Delta x}\big)  \right)=0. \label{neu3}
\end{align}
Eqs.~\eqref{neu2} and \eqref{neu3} can be rewritten in a matrix form as well such as $\mathbf M \cdot \mathbf f = \mathbf 0$ with $\mathbf f = (T_0, q_0)$, and thus $\det(\mathbf{M})=0$ provides a characteristic polynomial for $\xi$, $p(\xi)$,
\begin{align}
p(\xi)=&\frac{e^{-\iu \Delta x k}}{2 \Delta t^2 \Delta x^2 \xi ^2} \Big(-\Delta t^2 e^{-\iu \Delta x k} \left(e^{\iu \Delta x k}-1\right)^2 \left(a e^{\iu \Delta x j k} \xi ^n+a e^{\iu \Delta x (1+j) k} \xi ^n+2 e^{\iu \Delta x k} \left(\lambda  \xi ^2+a (\xi-1 ) \xi ^n\right)\right) \nonumber \\
&+2  \rho  c \Delta t \Delta x^2 e^{\iu \Delta x k} (\xi-1 ) \xi  +2 \rho  c \tau\Delta x^2 e^{\iu \Delta x k} (\xi -1)^2   \Big). \label{neu4}
\end{align}
It is worth noting that there are terms with $\xi^n$, i.e., it suggests that the stability properties may depend on the actual time step. However, the stability condition means $|\xi| \leq 1$, so that the scheme is meaningful only when it leads to stable solutions. When it does, then $\xi^n \rightarrow 0$, so that the remaining part of $p(\xi)$ determines the stability properties, and therefore must provide $|\xi| \leq 1$ automatically. In order to prove it, we use the Jury criteria, i.e., 
\begin{enumerate}
\item $p(\xi=1)\geq 0$;
\item $p(\xi=-1)\geq 0$;
\item $|a_0| \leq a_m$;
\end{enumerate}
for a polynomial in the form of $p(\xi) = a_m \xi^m + \dots + a_0$. Indeed, with the terms $\xi^n \rightarrow 0$, Eq.~\eqref{neu4} simplifies to a polynomial with coefficients of
\begin{align}
a_0 = \frac{\rho c \tau }{\Delta t^2 }, \quad a_1 = -\rho  c\frac{  {\Delta t}+2 \tau}{ {\Delta t}^2}, \quad  a_2 = \frac{ -  2\lambda \big( \cos{ (k \Delta x )}-1\big) {\Delta t}^2  + {\Delta t}  {\Delta x}^2 \rho  c+ {\Delta x}^2 { \rho  c} \tau }{ {\Delta t}^2  {\Delta x}^2}.
\end{align}
Since $-1\leq \cos(k \Delta x) \leq 1$, thus both situations must be checked. Straightforward calculations show that all enumerated conditions are automatically satisfied, so that the assumption $|\xi| \leq 1$ is valid, and $\xi^n \rightarrow 0$ indeed.

\begin{figure}
\centering
\includegraphics[width=15cm,height=6cm]{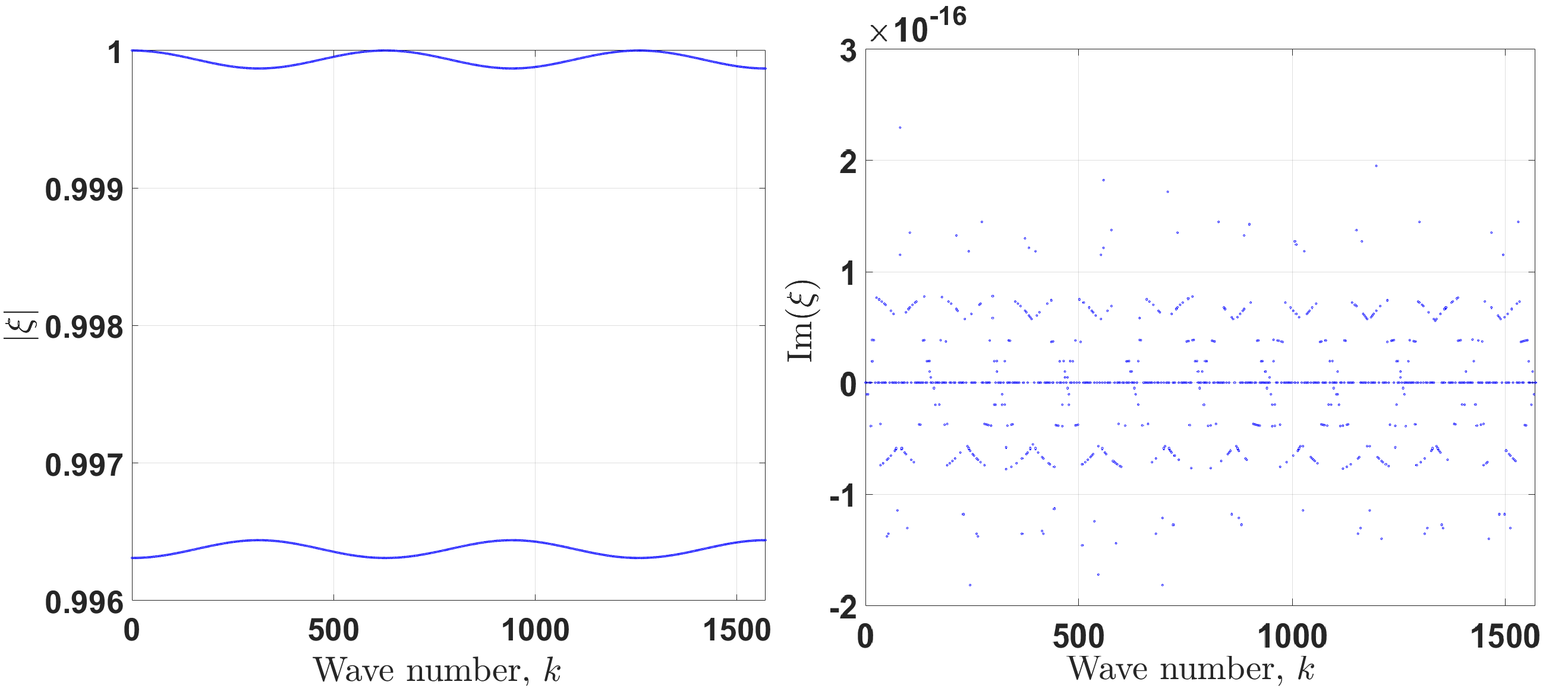}
\caption{The behavior of both roots as a function of the wave number. Left: the absolute value of the wave amplitude $\xi$. Right: The imaginary part of $\xi$. The wave number reaches $500 \pi$.}
\label{fig:FIG}
\end{figure}

The numerical dissipation is also characterized by $\xi$. If $|\xi|=1$, then the scheme is called conservative, free from dissipation errors, otherwise the scheme is called dissipative. The dispersion error is strongly related to the imaginary part of $\xi$. For a more detailed numerical characterization of such artificial errors, we refer to \cite{FulEtal20}. For the present nonlinear scheme, we can numerically investigate Eq.~\eqref{neu4}, and study its absolute value and imaginary parts. For this reason, let us assign the following values for the coefficients, so that $\rho = 2.5\cdot 10^{3}$ kg/m$^3$, $c = 700$ J/(kg K), $\tau = 0.27$ s, $\lambda = 5.5$ W/(m K), $a=2$ W/(m K$^2$). For the relaxation time, we used \cite{FehEtal21} for a realistic value for a rock material; and let $\Delta x=0.01$ m, $\Delta t=0.001$ s.
Fig.~\ref{fig:FIG} shows the behavior of the wave amplitude, being close to $1$, viz., conservative. Moreover, its imaginary part is practically zero over the entire region (such order of magnitude can also emerge from numerical errors of the root finding procedure), so that we do not expect a dispersion error as well. 

\section{Numerical simulation}
\subsection{Numerical simulation: Boundary type I.}
\begin{figure}[H]
\centering
\begin{subfigure}{0.38\textwidth}
\includegraphics[width=\linewidth]{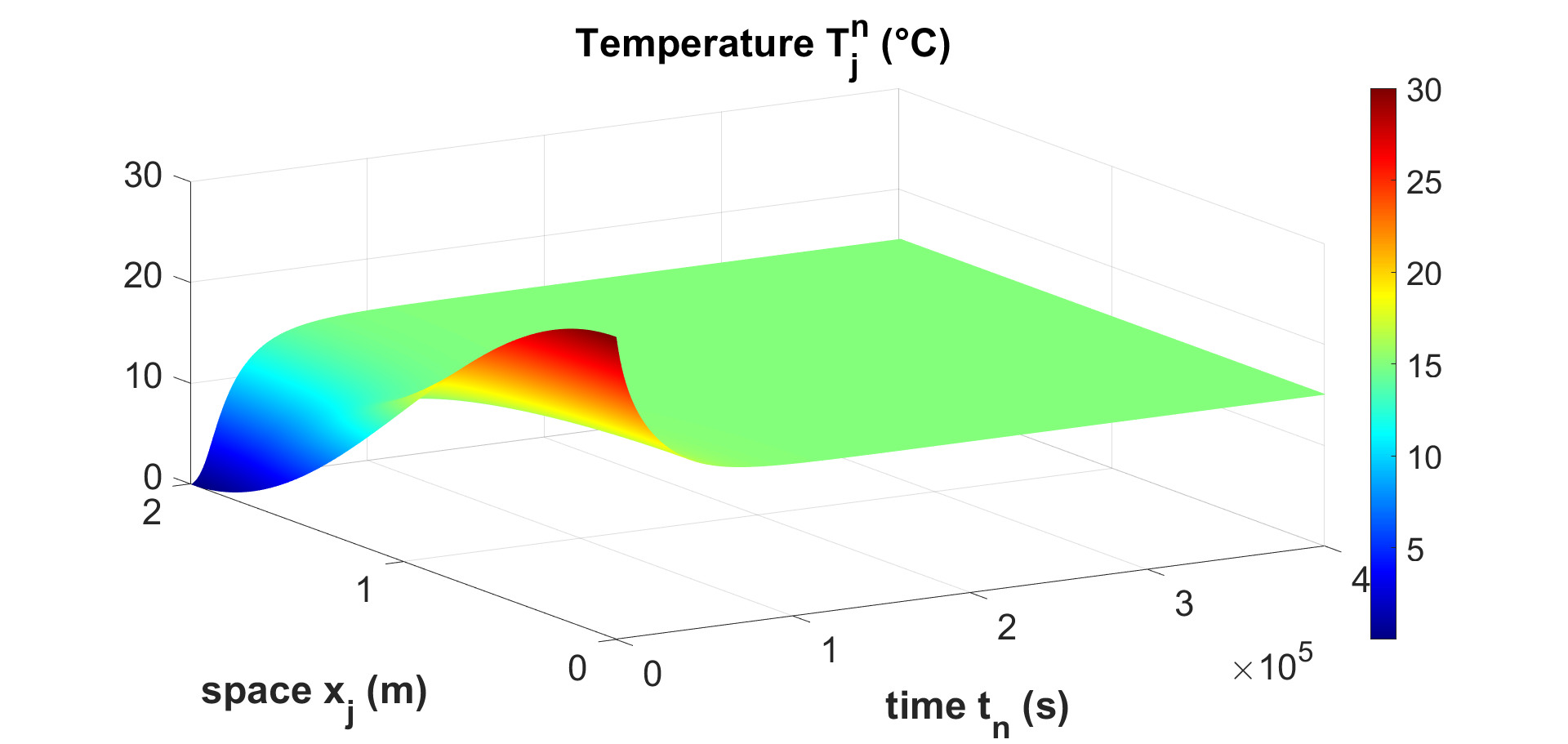}
\caption{Oblique view} \label{fig:T_nonlinear_01}
\end{subfigure}
\begin{subfigure}{0.38\textwidth}
\includegraphics[width=\linewidth]{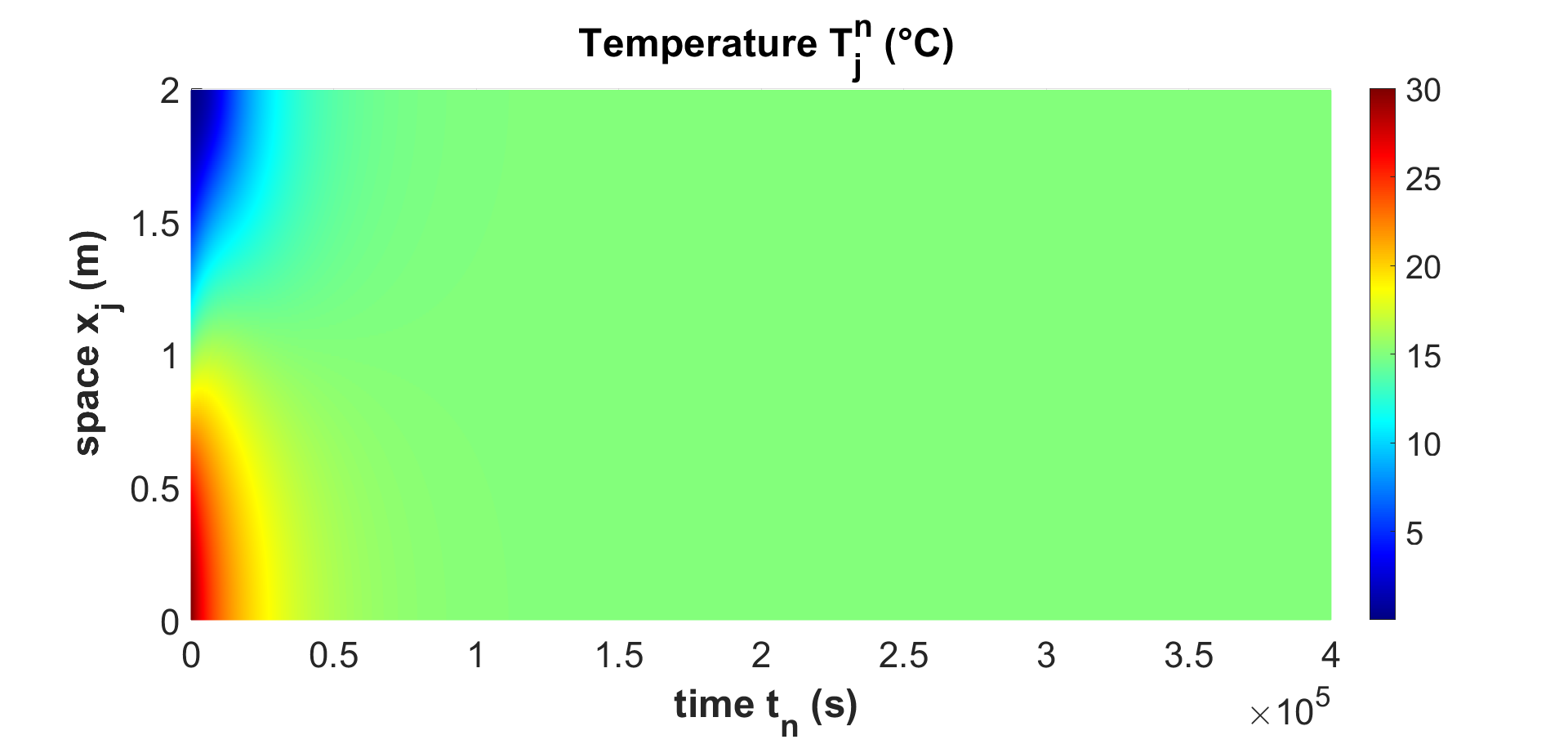}
\caption{Vertical view} \label{fig:T_nonlinear_02}
\end{subfigure}
\begin{subfigure}{0.38\textwidth}
\includegraphics[width=\linewidth]{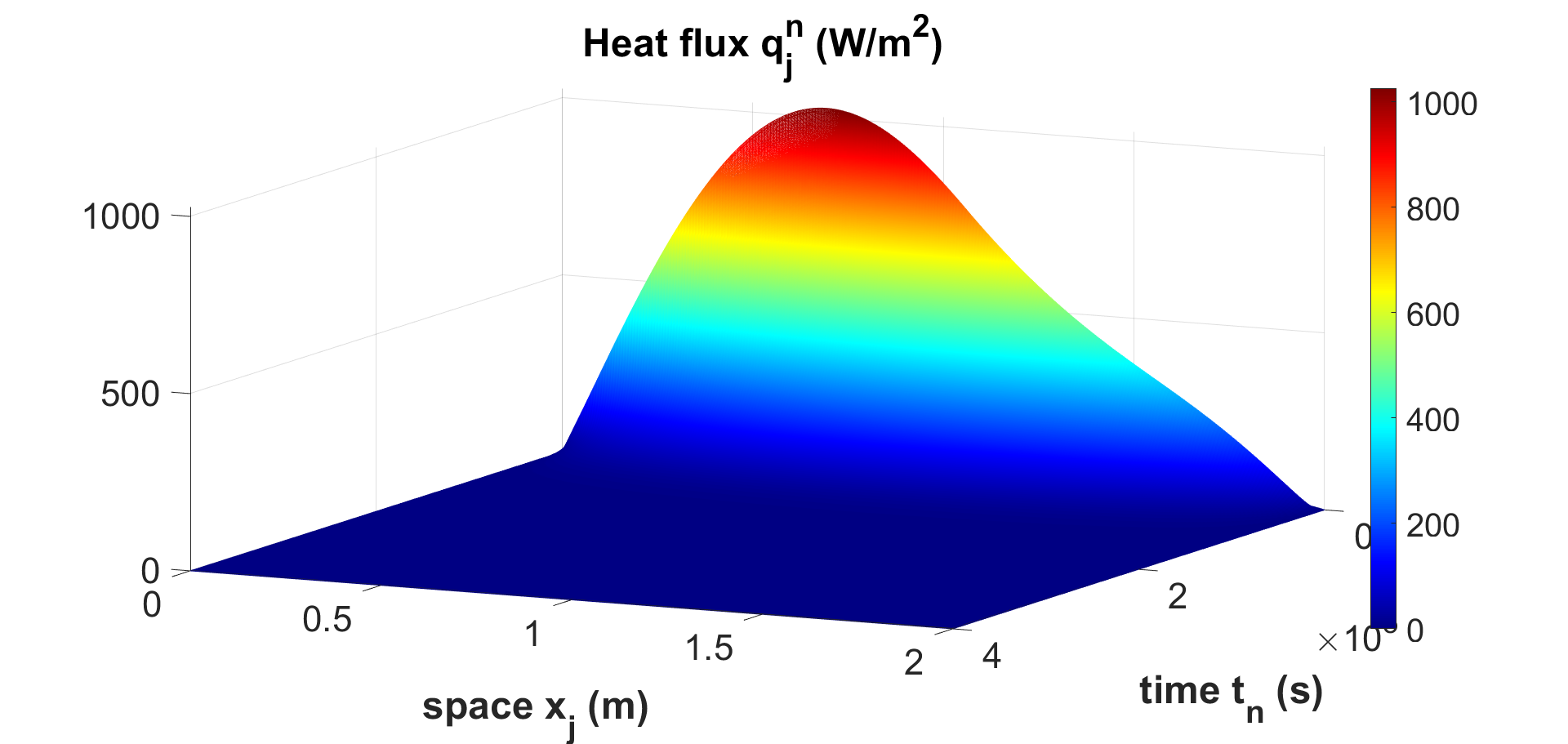}
\caption{Oblique view} \label{fig:Q_nonlinear_01}
\end{subfigure}
\begin{subfigure}{0.38\textwidth}
\includegraphics[width=\linewidth]{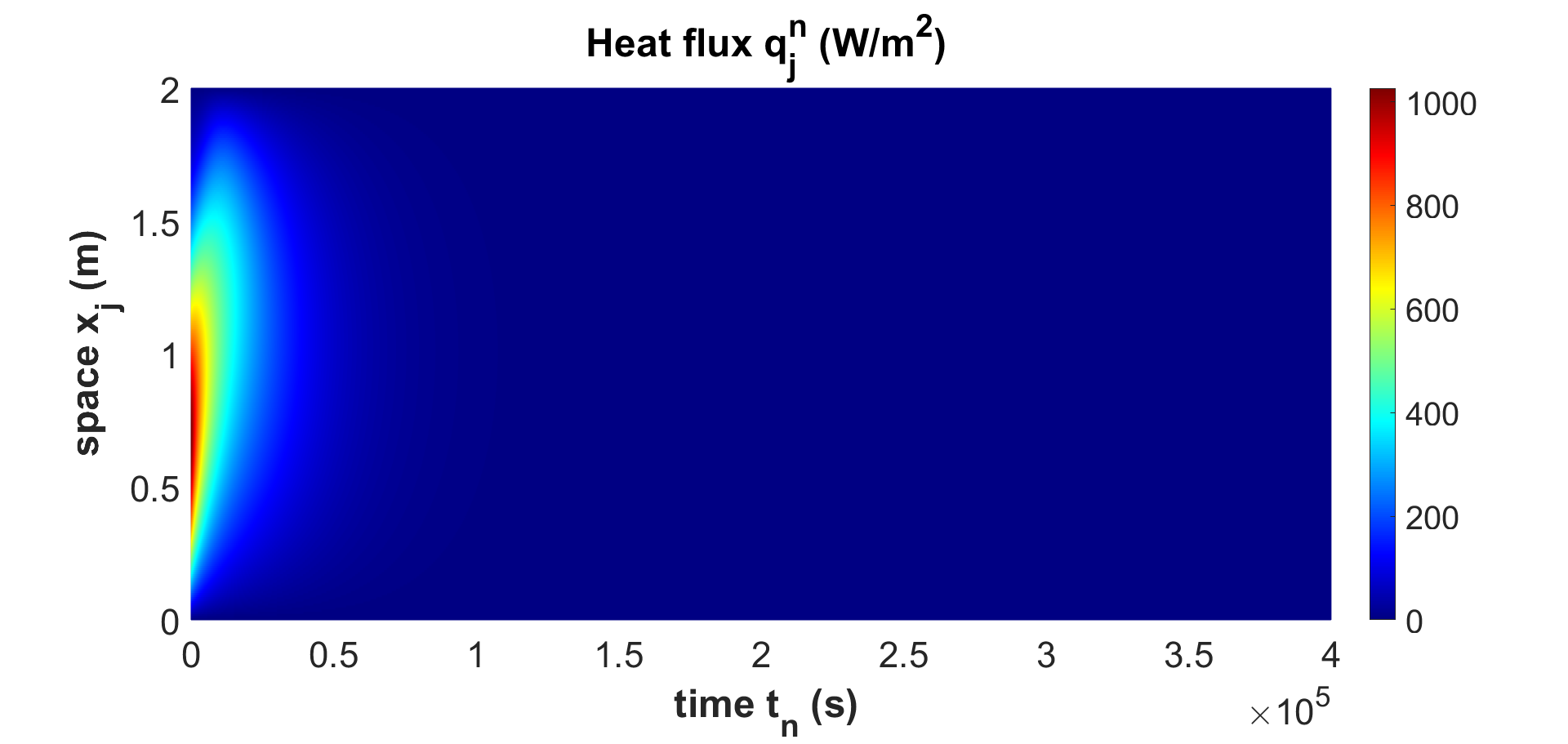}
\caption{Vertical view} \label{fig:Q_nonlinear_02}
\end{subfigure}
\caption{Simulation results for boundary type I.~for the nonlinear case with $a = 2 $ W/(m K$^2$). Figures (a, b) demonstrate the time evolution of the temperature field. Figure (c,d) show the corresponding heat flux field.} 
\label{fig:1a}
\end{figure}

\begin{figure}[H]
\centering
\begin{subfigure}{0.38\textwidth}
\includegraphics[width=\linewidth]{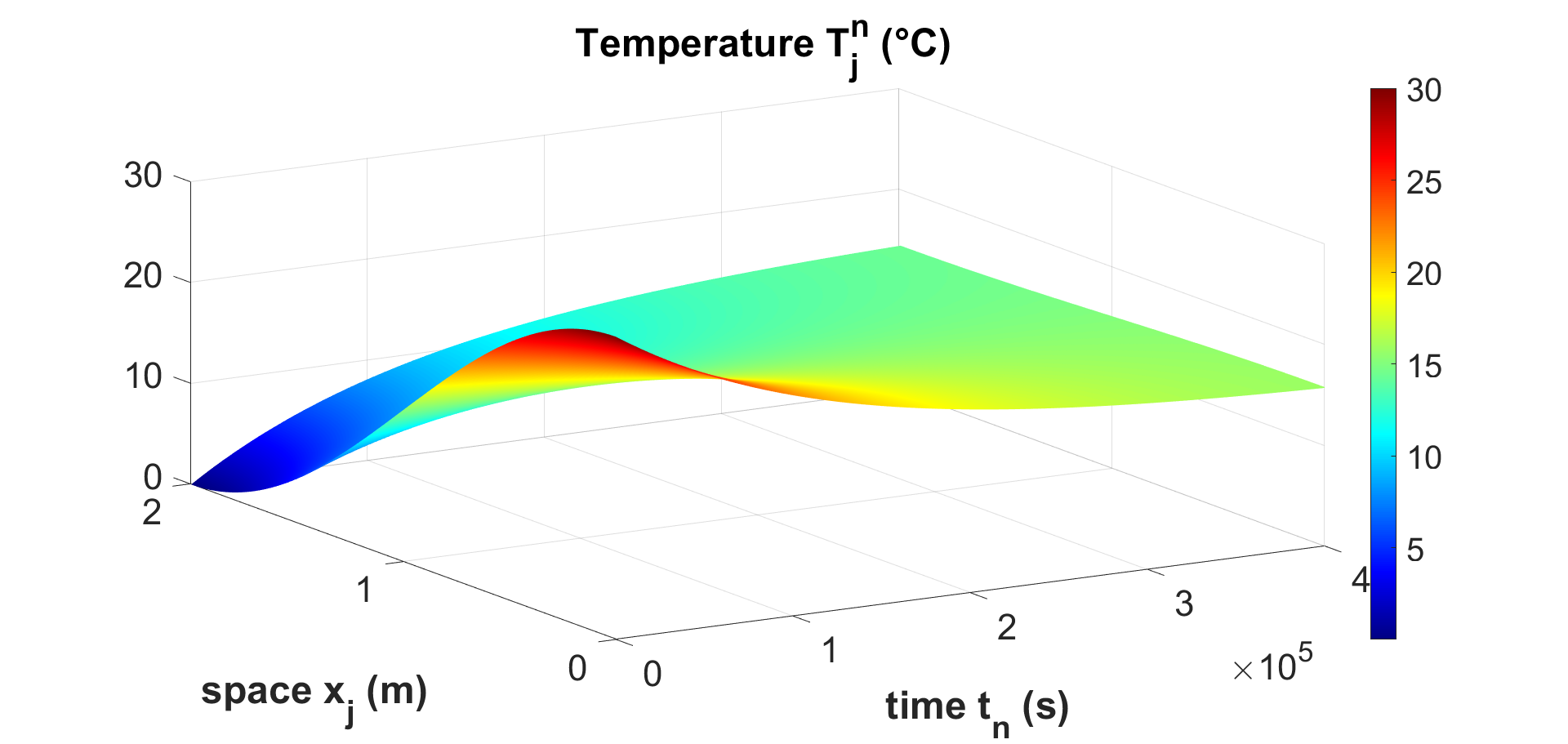}
\caption{Oblique view} \label{fig:T_linear_01}
\end{subfigure}
\begin{subfigure}{0.38\textwidth}
\includegraphics[width=\linewidth]{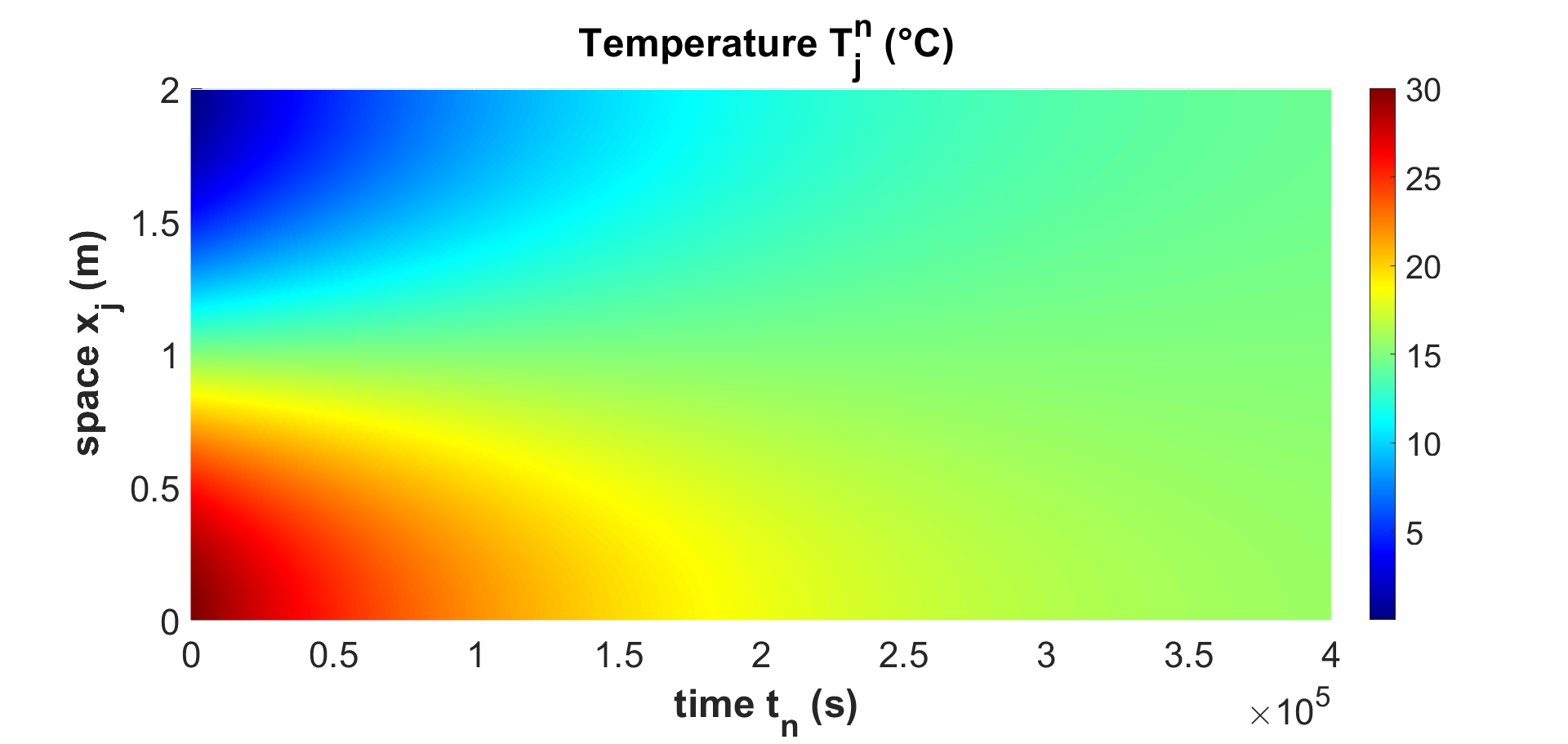}
\caption{Vertical view} \label{fig:T_linear_02}
\end{subfigure}
\begin{subfigure}{0.38\textwidth}
\includegraphics[width=\linewidth]{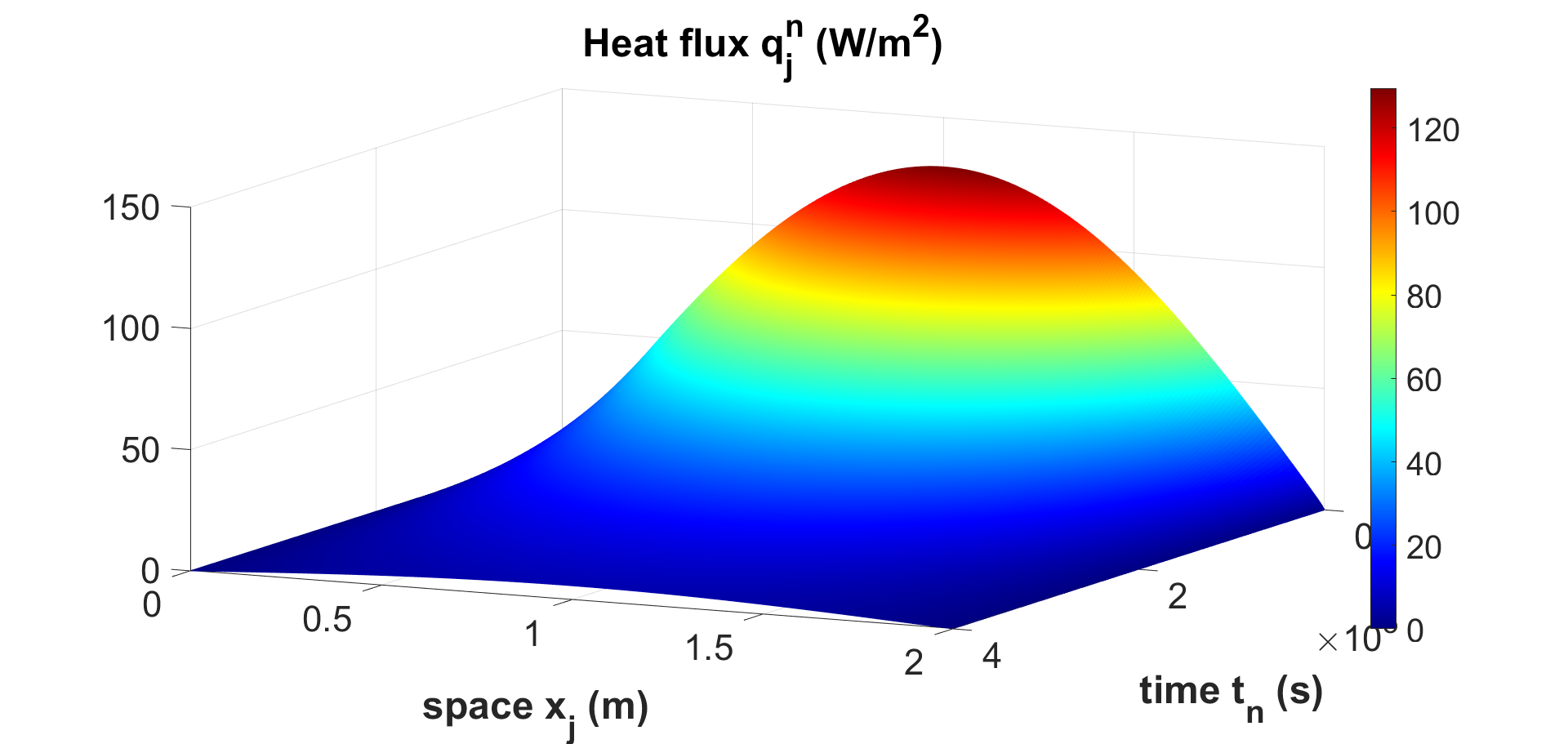}
\caption{Oblique view} \label{fig:Q_linear_01}
\end{subfigure}
\begin{subfigure}{0.38\textwidth}
\includegraphics[width=\linewidth]{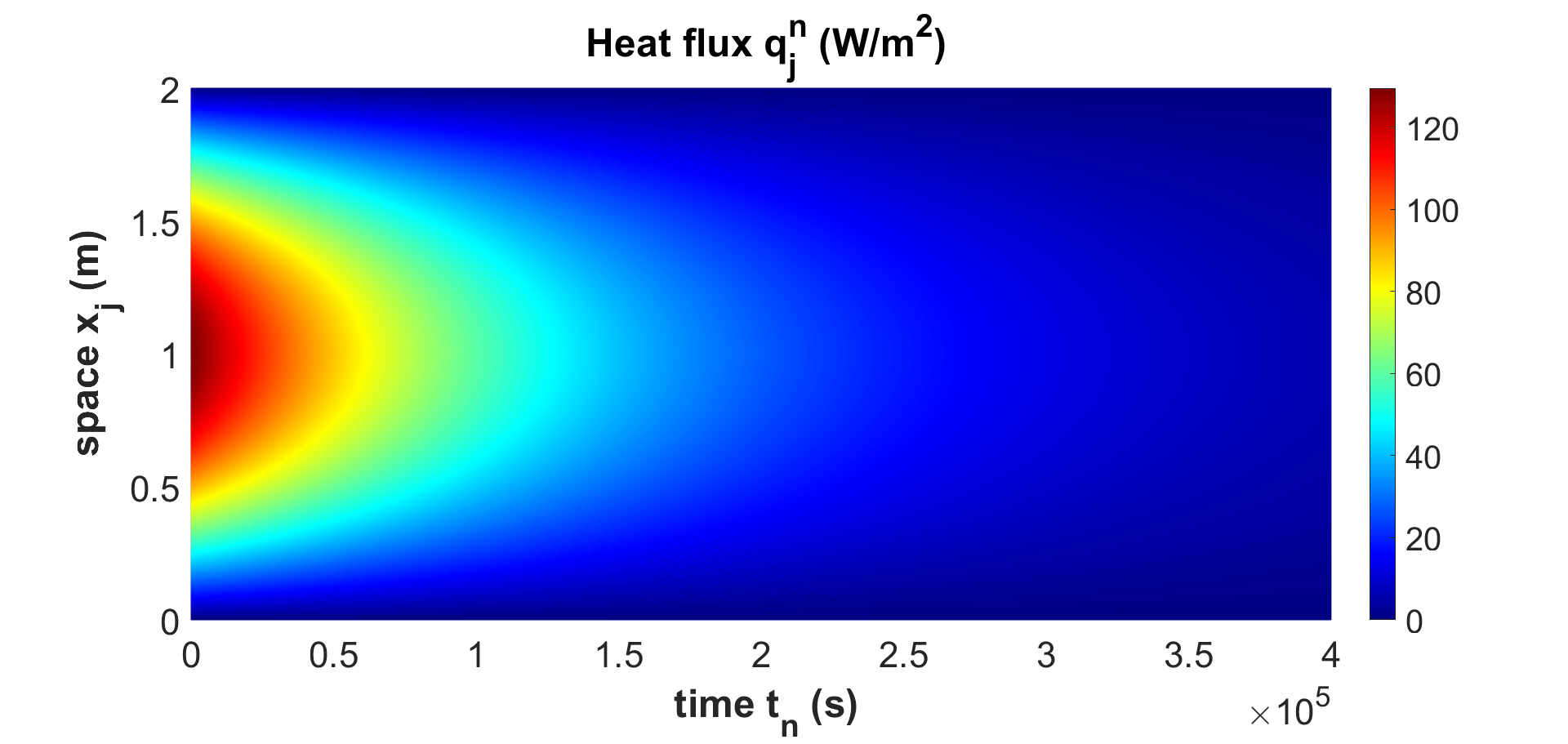}
\caption{Vertical view} \label{fig:Q_linear_02}
\end{subfigure}
\caption{Simulation results for boundary type I.~for the linear case with $a = 0 $ W/(m K$^2$). Figures (a, b) demonstrate the time evolution of the temperature field. Figure (c,d) show the corresponding heat flux field.} 
\label{fig:2a}
\end{figure}

In this section, we implement the same parameters and solve the difference equations for both types of boundary conditions as a brief demonstration, for both linear ($a=0$) and nonlinear ($a>0$) situations.
We  use the following initial conditions associated to the discrete nonlinear system \eqref{syst.01}
\begin{eqnarray}\label{initial.conditions}
T_{j}^{0}=T_{b}+\frac{T_{f}}{2}\cos\Big(\frac{\pi x_{j}}{\ell}\Big), 
\quad 
{q_{j}^{0}=\frac{a\pi T_{f}}{2\ell}\bigg[T_{b}+\frac{T_{f}}{2}\cos\Big(\frac{\pi x_{j}}{\ell}\Big)\bigg]\sin\Big(\frac{\pi x_{j}}{\ell}\Big)+\frac{\lambda\pi T_{f}}{2\ell}\sin\Big(\frac{\pi x_{j}}{\ell}\Big)},  \quad {j=0, 1,...,J},
\end{eqnarray}
where let $T_{b}=15$ $^\circ$C and $T_{f}=30$ $^\circ$C, so that initial temperature distribution can be realistic for a practical situation. We also want to highlight here that the initial heat flux field is determined in agreement with the nonlinear constitutive relation in order to avoid incompatibility and nonphysical behavior. Figures \ref{fig:1a} and \ref{fig:2a} demonstrates the characteristic differences between the nonlinear and linear cases. It is apparent that due to the temperature dependent thermal conductivity, the initial heat flux field and its time evolution cannot be symmetric.

\subsection{Numerical simulation: Boundary type II}

\begin{figure}[H]
\centering
\begin{subfigure}{0.45\textwidth}
\includegraphics[width=8cm,height=5.3cm]{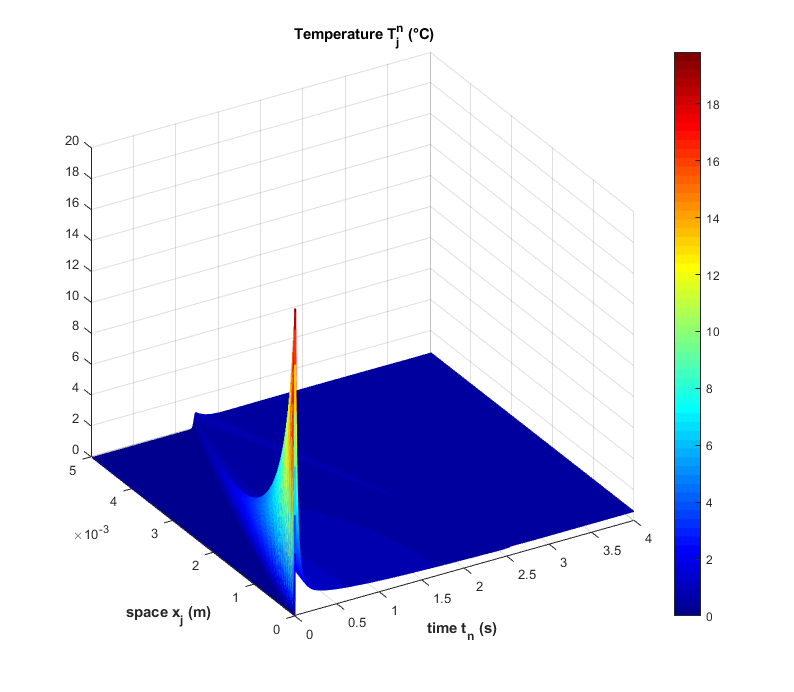}
\caption{Oblique view} \label{fig:T_nonlinear_01_tp}
\end{subfigure}
\begin{subfigure}{0.45\textwidth}
\includegraphics[width=8cm,height=5.3cm]{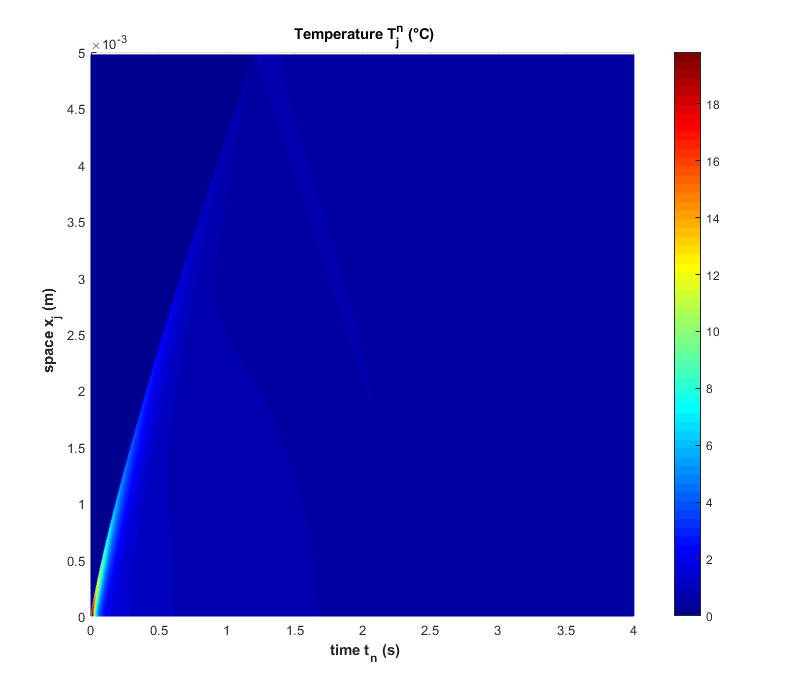}
\caption{Vertical view} \label{fig:T_nonlinear_02_tp}
\end{subfigure}
\begin{subfigure}{0.45\textwidth}
\includegraphics[width=8cm,height=5.3cm]{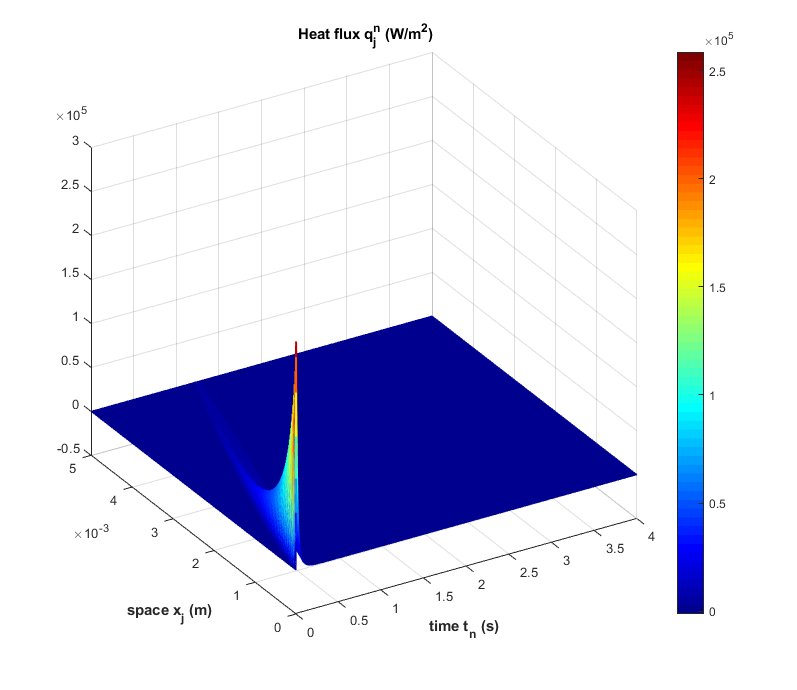}
\caption{Oblique view} \label{fig:Q_nonlinear_01_tp}
\end{subfigure}
\begin{subfigure}{0.45\textwidth}
\includegraphics[width=8cm,height=5.3cm]{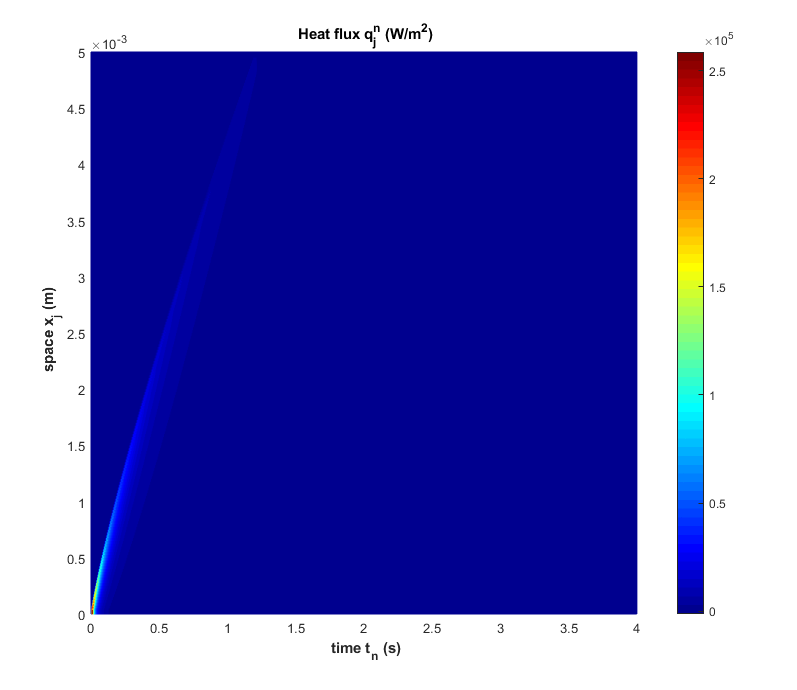}
\caption{Vertical view} \label{fig:Q_nonlinear_02_tp}
\end{subfigure}
\caption{Simulation results for boundary type II.~for the nonlinear case with $a = 2 $ W/(m K$^2$). Figures (a, b) demonstrate the time evolution of the temperature field. Figure (c,d) show the corresponding heat flux field.} 
\label{fig:3a}
\end{figure}

\begin{figure}[H]
\centering
\begin{subfigure}{0.45\textwidth}
\includegraphics[width=8cm,height=5.3cm]{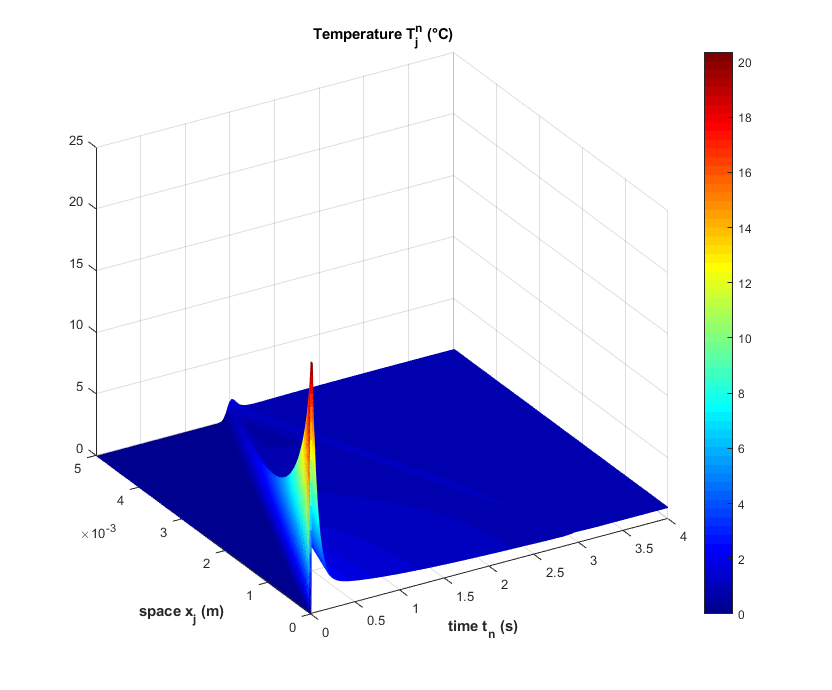}
\caption{Oblique view} \label{fig:T_linear_01_tp}
\end{subfigure}
\begin{subfigure}{0.45\textwidth}
\includegraphics[width=8cm,height=5.3cm]{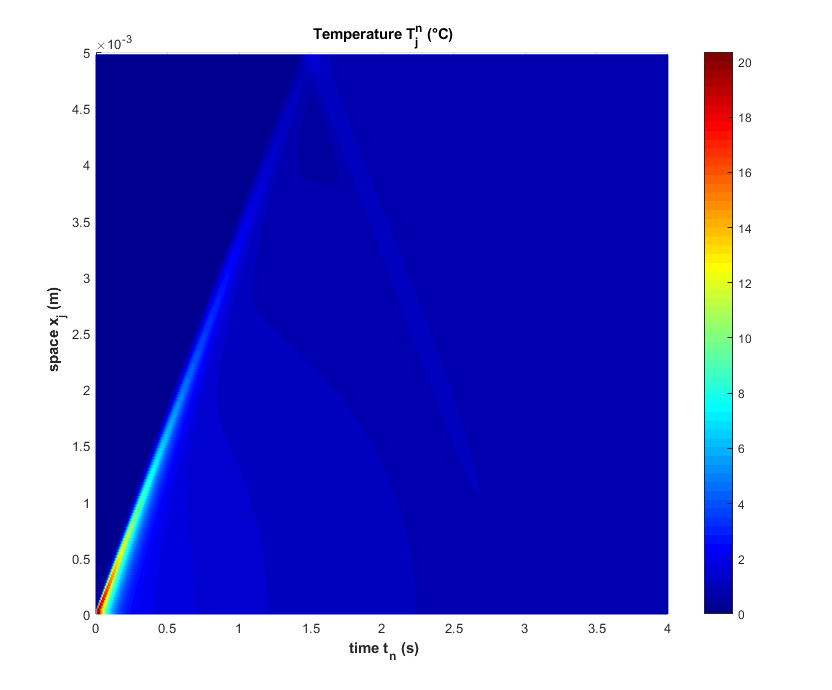}
\caption{Vertical view} \label{fig:T_linear_02_tp}
\end{subfigure}
\begin{subfigure}{0.45\textwidth}
\includegraphics[width=8cm,height=5.3cm]{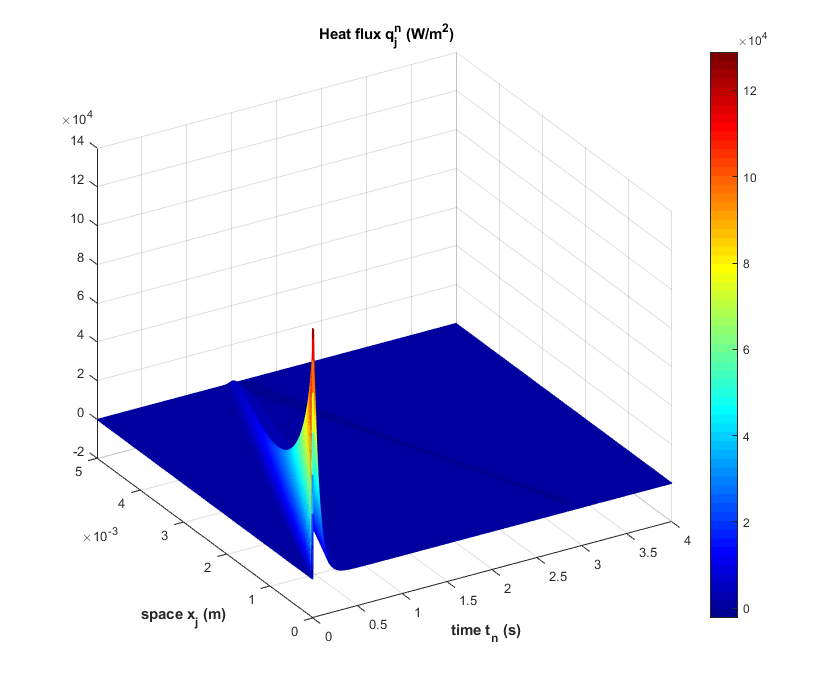}
\caption{Oblique view} \label{fig:Q_linear_01_tp}
\end{subfigure}
\begin{subfigure}{0.45\textwidth}
\includegraphics[width=8cm,height=5.3cm]{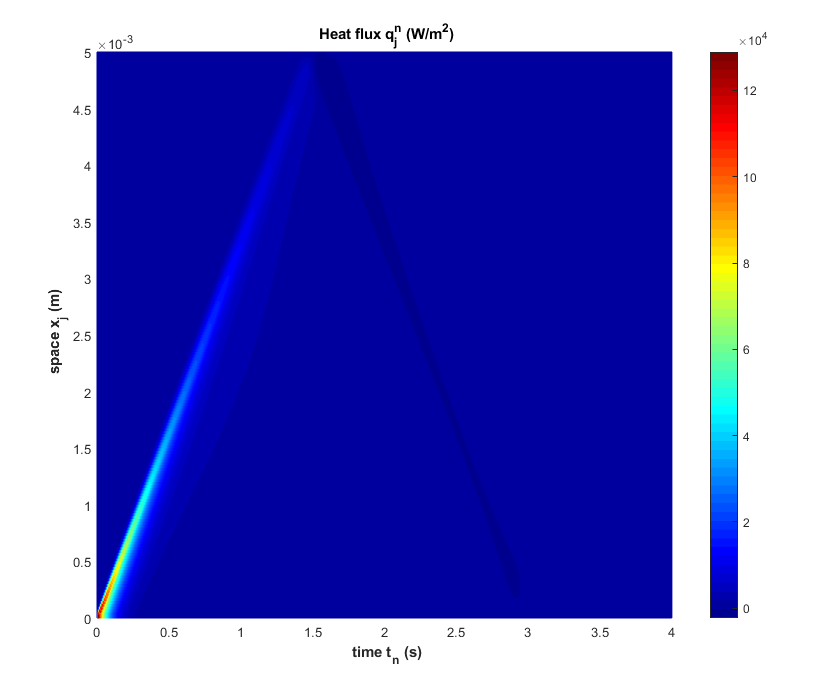}
\caption{Vertical view} \label{fig:Q_linear_02_tp}
\end{subfigure}
\caption{Simulation results for boundary type II.~for the linear case with $a = 0 $ W/(m K$^2$). Figures (a, b) demonstrate the time evolution of the temperature field. Figure (c,d) show the corresponding heat flux field.} 
\label{fig:4a}
\end{figure}

We  use the following initial conditions associated to the discrete system \eqref{syst.01}
\begin{eqnarray}\label{initial.conditions2}
T_{j}^{0}=0, \quad q_{j}^{0}=0,  \quad {j=0, 1,...,J},
\end{eqnarray}
with the same parameters as previously.
Below are the simulations for the nonlinear and linear cases. Figures \ref{fig:3a} and \ref{fig:4a} present the corresponding time evolution temperature and heat flux fields for time-dependent boundary condition, for both linear and nonlinear cases.

\begin{figure}[H]
\centering
\includegraphics[width=9.5cm,height=7cm]{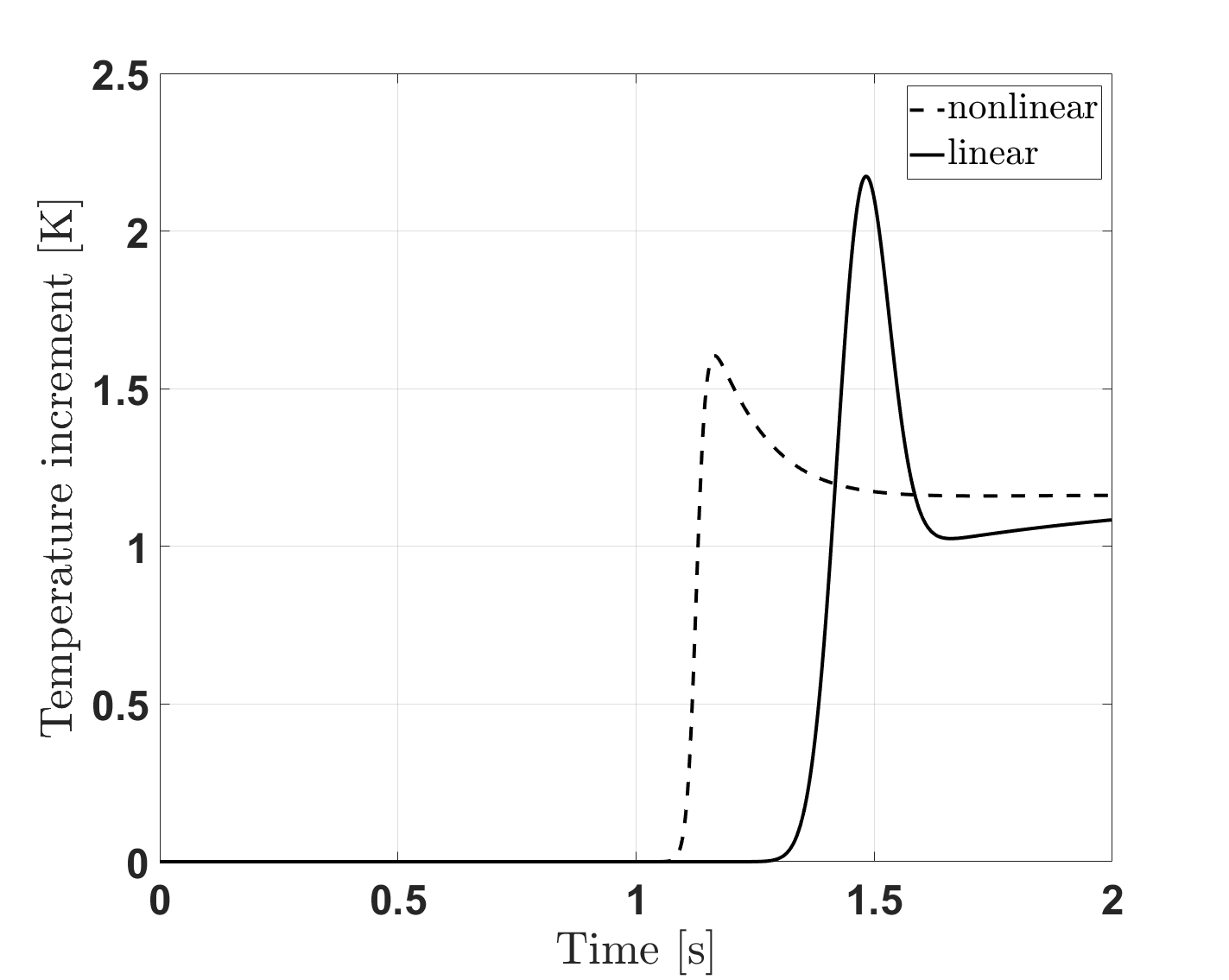}
\caption{Demonstrating the differences between the linear and nonlinear cases for $a=0$ W/(m K$^2$) and $a=2$ W/(m K$^2$). }
\label{fig:rs}
\end{figure}

Since the rear side temperature history has of greater practical importance in case of this settings, we also compare the linear and nonlinear solutions for boundary type II, see Figure \ref{fig:rs} for details. It reveals that such a nonlinear behavior can significantly distort the wave signal. Furthermore, as the thermal conductivity is progressively behaves with respect to temperature, the propagation speed of the wave front becomes higher than in the linear case, this expectation is also apparent in Fig.~\ref{fig:rs}. Furthermore, the wave amplitude decreases since the thermal diffusivity becomes larger, thus it notably dampens that wave front. The steady-state, however, must not change with identical heat capacities as the boundaries are adiabatic and the heat pulse provides the same energy.

\section{Summary}

In the present paper, we have investigated the transient behavior of the nonlinear Cattaneo equation, including a temperature-dependent thermal conductivity. Additionally, we have proposed an implicit numerical scheme, which is unconditionally stable and also free from numerical dispersion. Such scheme has enabled to use relatively steep temperature-dependence in thermal conductivity without introducing significant artificial distortion into the numerical solution. 

For demonstration, we solved two different settings. In the first one, we simulated the evolution of an inhomogeneous initial temperature distribution. The correct setting has required the determination of a compatible initial heat flux field. It is clear how the nonlinearity distorts the symmetry. However, such initial state cannot reflect the influence of nonlinearities on wave propagation. For this reason, we also included a more practical setting in the second simulation, using a heat pulse boundary condition. Since such experiment is used to determine the material properties based on the temperature history, we also demonstrated the effects of such nonlinearity on the rear side temperature evolution. The simulation shows that the wave front can be considerably damped by the increasing thermal diffusivity, however, the front becomes faster as well. Therefore the simulations are physically sound, and the present numerical scheme provides a basis for future research.

\section*{Funding}

Project no.~TKP-6-6/PALY-2021 has been implemented with the support provided by the Ministry of Culture and Innovation of Hungary from the National Research, Development and Innovation Fund, financed under the TKP2021-NVA funding scheme. The research was funded by the János Bolyai Research Scholarship of the Hungarian Academy of Sciences, and by the National Research, Development and Innovation Office-NKFIH FK 134277.

\section*{Declarations}
\textbf{Conflict of interest} The author declares no competing interests.

\section*{References}
\bibliographystyle{unsrt}

\begin{thebibliography}{10}
	\expandafter\ifx\csname url\endcsname\relax
	\def\url#1{\texttt{#1}}\fi
	\expandafter\ifx\csname urlprefix\endcsname\relax\def\urlprefix{URL }\fi
	\expandafter\ifx\csname href\endcsname\relax
	\def\href#1#2{#2} \def\path#1{#1}\fi
	
	\bibitem{JacWalMcN70}
	H.~E. Jackson, C.~T. Walker, T.~F. McNelly, Second sound in {N}a{F}, Physical
	Review Letters 25~(1) (1970) 26--28.
	
	\bibitem{Don09}
	R.~J. Donnelly, The two-fluid theory and second sound in liquid {H}elium,
	Physics Today 62~(10) (2009) 34--39.
	
	\bibitem{Struc05}
	H.~Struchtrup, Macroscopic {T}ransport {E}quations for {R}arefied {G}as
	{F}lows, Springer, 2005.
	
	\bibitem{Arietal15}
	T.~Arima, T.~Ruggeri, M.~Sugiyama, S.~Taniguchi, Non-linear extended
	thermodynamics of real gases with 6 fields, International Journal of
	Non-Linear Mechanics 72 (2015) 6--15.
	
	\bibitem{RhaStruc16}
	B.~Rahimi, H.~Struchtrup, Macroscopic and kinetic modelling of rarefied
	polyatomic gases, Journal of Fluid Mechanics 806 (2016) 437--505.
	
	\bibitem{SellEtal16b}
	A.~Sellitto, V.~A. Cimmelli, D.~Jou, Mesoscopic theories of heat transport in
	nanosystems, Springer, Berlin, 2016.
	
	\bibitem{Chen00}
	G.~Chen, Phonon heat conduction in nanostructures, International Journal of
	Thermal Sciences 39~(4) (2000) 471--480.
	
	\bibitem{FehEtal21}
	A.~Fehér, N.~Lukács, L.~Somlai, T.~Fodor, M.~Szücs, T.~Fülöp, P.~Ván,
	R.~Kovács, Size effects and beyond-{F}ourier heat conduction in
	room-temperature experiments, Journal of Non-Equilibrium Thermodynamics 46
	(2021) 403--411.
	
	\bibitem{LunEtal22}
	A.~Lunev, A.~Lauerer, V.~Zborovskii, F.~Léonard, Digital twin of a laser flash
	experiment helps to assess the thermal performance of metal foams,
	International Journal of Thermal Sciences 181 (2022) 107743.
	
	\bibitem{MulRug98}
	I.~Müller, T.~Ruggeri, Rational Extended Thermodynamics, Springer, 1998.
	
	\bibitem{RugSug15}
	T.~Ruggeri, M.~Sugiyama, Rational extended thermodynamics beyond the monatomic
	gas, Springer, 2015.
	
	\bibitem{JouEtal10b}
	D.~Jou, J.~Casas-Vázquez, G.~Lebon, Extended {I}rreversible {T}hermodynamics,
	Springer Verlag, Berlin, 2010 (fourth Edition), 2010.
	
	\bibitem{Lebon14}
	G.~Lebon, Heat conduction at micro and nanoscales: a review through the prism
	of extended irreversible thermodynamics, Journal of Non-Equilibrium
	Thermodynamics 39~(1) (2014) 35--59.
	
	\bibitem{Gyarmati70b}
	I.~Gyarmati, Non-equilibrium thermodynamics, Springer, 1970.
	
	\bibitem{Verhas96}
	J.~Verh{\'a}s, Once again on the transport of dynamic degrees of freedom, Atti
	Accademia Peloritana dei Pericolanti 72 (1996) 101--114.
	
	\bibitem{BerVan17b}
	A.~Berezovski, P.~V\'an, Internal Variables in Thermoelasticity, Springer,
	2017.
	\newblock \href {http://dx.doi.org/10.1007/978-3-319-56934-5}
	{\path{doi:10.1007/978-3-319-56934-5}}.
	
	\bibitem{VanFul12}
	P.~Ván, T.~Fülöp, Universality in heat conduction theory -- weakly nonlocal
	thermodynamics, Annalen der Physik (Berlin) 524~(8) (2012) 470--478.
	
	\bibitem{Cattaneo58}
	C.~Cattaneo, Sur une forme de lequation de la chaleur eliminant le paradoxe
	dune propagation instantanee, Comptes Rendus Hebdomadaires Des Seances De
	L'Academie Des Sciences 247~(4) (1958) 431--433.
	
	\bibitem{Vernotte58}
	P.~Vernotte, Les paradoxes de la th{\'e}orie continue de l{\'e}quation de la
	chaleur, Comptes Rendus Hebdomadaires Des Seances De L'Academie Des Sciences
	246~(22) (1958) 3154--3155.
	
	\bibitem{Auriault16}
	J.-L. Auriault, Cattaneo–{V}ernotte equation versus {F}ourier thermoelastic
	hyperbolic heat equation, International Journal of Engineering Science 101
	(2016) 45--49.
	\newblock \href
	{http://dx.doi.org/https://doi.org/10.1016/j.ijengsci.2015.12.002}
	{\path{doi:https://doi.org/10.1016/j.ijengsci.2015.12.002}}.
	
	\bibitem{KovRog20}
	R.~Kov{\'a}cs, P.~Rogolino, Numerical treatment of nonlinear {F}ourier and
	{M}axwell-{C}attaneo-{V}ernotte heat transport equations, International
	Journal of Heat and Mass Transfer 150 (2020) 119281.
	
	\bibitem{Press07b}
	W.~H. Press, Numerical recipes 3rd edition: {T}he art of scientific computing,
	Cambridge University Press, 2007.
	
	\bibitem{FulEtal20}
	T.~Fülöp, R.~Kovács, M.~Szücs, M.~Fawaier, Thermodynamical extension of a
	symplectic numerical scheme with half space and time shifts demonstrated on
	rheological waves in solids, Entropy 22 (2020) 155.
	
\end{thebibliography}

\end{document}